\documentclass{elsart}
\usepackage{graphicx,latexsym,amssymb,natbib}
\newcommand {\sax} {{\it Beppo}SAX }
\newcommand {\hess} {H.E.S.S. }

\newcommand {\C}{\v Cerenkov }

\newcommand {\gamm} {$\gamma$}

\begin{document}
%\runauthor{Cicero, Caesar and Vergil}

\begin{frontmatter}
\title{A brief (blazar oriented) overview on topics for multi-wavelength observations with \\ TeV photons}
%\title{A brief (blazar oriented) scientific overview on multi-wavelength observations with TeV photons}

%\vspace*{-0.6cm}
\author[mpik]{L. Costamante}
\address[mpik]{Max-Planck Institut f\"ur Kernphysik, Heidelberg, Germany}

%\vspace*{-0.4cm}
\begin{abstract}
Multi-wavelength observations with TeV photons are an essential diagnostic tool
to study the physics of TeV sources. The X-ray and optical bands are especially valuable, since
they sample TeV energy electrons and their most effective seed photons for the inverse Compton up-scattering.
The complex variability of blazars, however (timescales from years down to minutes, 
with different patterns and SED behaviours),
requires a great effort on simultaneous campaigns, which should be performed possibly over several days. 
Most important is the possibility to have spectra in all the three bands: with the new instruments, both in 
TeV and X-ray bands, this has become now feasible on timescales less than one hour.
The insights from such observations can be tremendous, since recent results have shown that the X-ray
and TeV emissions do not always follow the same behaviour, and  flares can have different relations between 
rise and decay times. Unfortunately, the strong pointing constraints of XMM do not allow the full use of this satellite
simultaneously with ground telescopes. 
\end{abstract}

\end{frontmatter}
%\vspace*{-1.2cm}

\section{Introduction}
\vspace*{-0.8cm}
With the upcoming of the new generation of \C Telescopes (CT, e.g. H.E.S.S., VERITAS, CANGAROO III, MAGIC),
TeV astronomy is about to greatly improve its scientific capabilities.
Multi-wavelength observations are an essential tool to investigate how TeV sources work.
Below, I will briefly present some recent topics and perspectives for multi-wavelength
observations with TeV photons, with  main emphasis on blazars since they represent nearly
all the extragalactic sources so far detected.
The scheme is the following: at first I will very briefly recall different mechanisms 
to produce TeV photons, to identify
the most interesting frequencies to accompany TeV observations. Then I will discuss
the objects' main variability characteristics, which determine the observation strategies,
and will focus on some issues related to the simultaneous coverage.

%\vspace*{-0.5cm}
\section{TeV emission ingredients}
\vspace*{-0.8cm}
The first obvious requirement to produce TeV photons is to have particles with TeV energies,
for energy conservation. According to which particle is the main carrier of this energy, 
%that will eventually go into the TeV emission, 
the recipes for TeV emission
%models developed to explain the TeV emission %\C telescopes (CT) data 
%mechanisms to explain the TeV emission 
can be divided in two classes. 
In the so called ``hadronic" models, protons play the main role,  producing TeV photons through
hadronic interactions (as $pp\to\pi^0\to2\gamma$ \cite{pohl}, or $p\gamma\to\pi\to\gamma e^+e^-$ with a
subsequent electromagnetic cascade \cite{mann}), or directly through synchrotron radiation 
in strong magnetic fields (B$\sim$100G), with emission at lower frequencies contributed also
by secondary pair-produced electrons \cite{ahap}. For  these models the overall 
spectral energy distribution (SED) is determined by the details of the cascade or the 
secondary electrons production,
as well as the initial particle spectrum and source conditions, so that basically all 
lower frequencies are equally interesting to distinguish among 
them and to constrain the physical parameters.

In the ``leptonic" models, instead, electrons are the main particles, producing TeV photons through
the inverse Compton (IC)  process (more economic than the synchrotron one). Allowing for beaming effects,
to give origin to an observed TeV photon (i.e. a TeV/$\delta$ photon in the comoving frame), 
an electron needs at least an energy $\gamma m_ec^2\geq {\rm TeV}/\delta$ (i.e. $\gamma\geq10^5$
for $\delta\sim10$). The seed photons  most effective to be up-scattered by these
electrons are of frequency $h\nu\leq m_ec^2/\gamma$ (in the co-moving frame), since above they
scatter in the Klein-Nishina regime. In the presence of magnetic field, these same electrons
will also produce synchrotron radiation of observed energy 
$h\nu = 1.5 B({\rm Gauss})(\gamma/10^5)^2(\delta/10)$ keV.
In this scenario, therefore, 
two other frequency bands stand out in a natural way: 
the X-ray one, which can provide number density and energy distribution
of the electrons also emitting in the TeV band, 
%thus providing a powerful diagnostic tool,
and the ``optical" one (in a broad sense, i.e. from UV to IR), where 
we can have information about the most effective seed photons.  
Multi-wavelength observations in these three bands represent therefore 
a powerful diagnostic tool for the acceleration mechanism and the 
details of the high energy emission (as well as a test for the leptonic scenario itself).

Note anyway that, whatever scenario is adopted,  
the conditions must be compatible
with the TeV photons survival  (since we detect them), i.e. the source must be transparent with respect to the
\gamm-\gamm~absorption.

%Note also that the overall production of photons in the source has to be compatible with the survival of TeV
%photons, since we want to see them (this is e.g. given by the use of beaming delta, 
%which lower the densities and requirements for emission) 

\begin{figure*}[t]
\centering
%\vspace*{-1.0cm}
\hspace*{-1.2cm}
\resizebox{17.7cm}{!}{\includegraphics[angle=0,width=10cm]{501test.epsi}
\hspace*{-0.2cm} \includegraphics[angle=0,width=12cm]{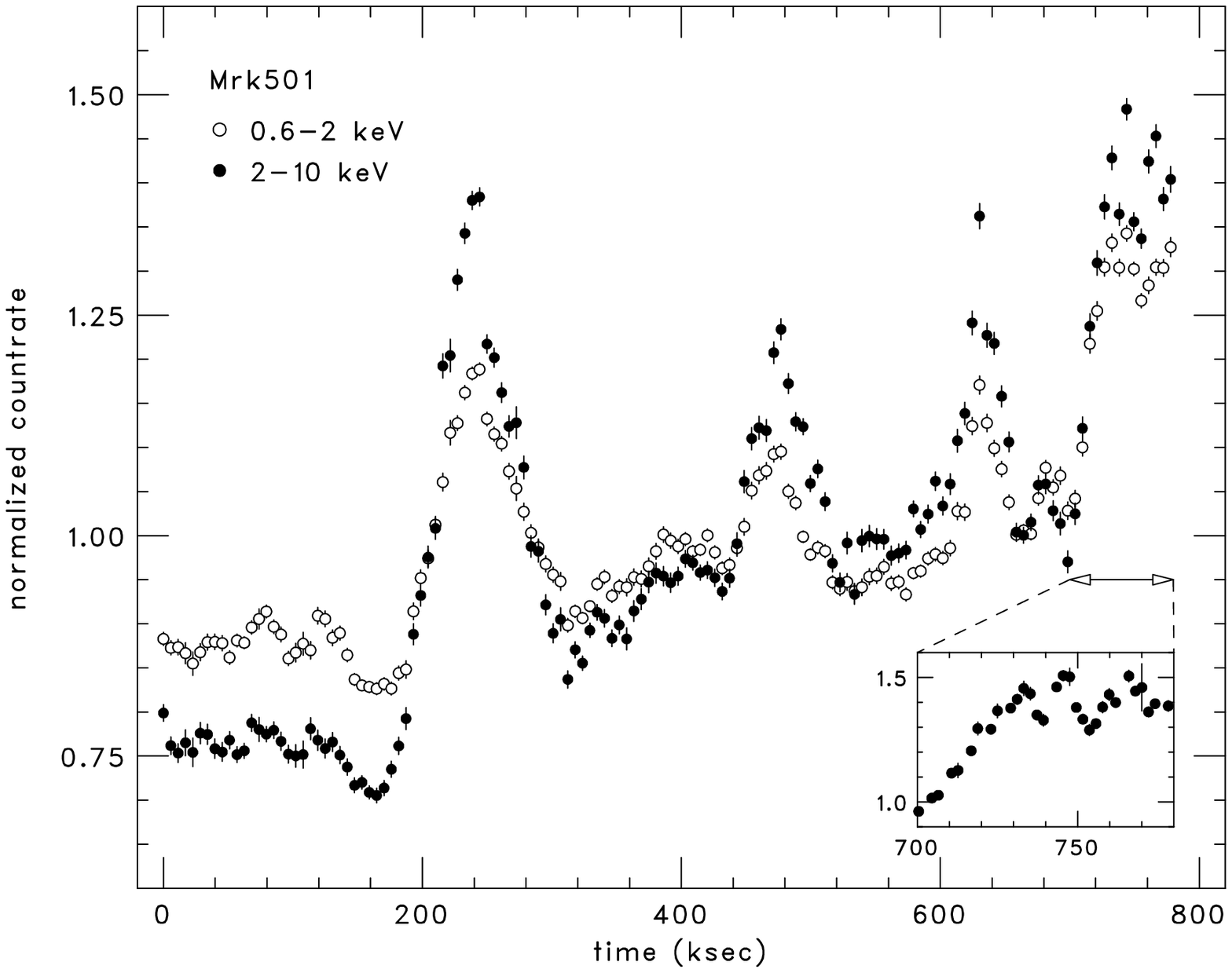}
\hspace*{-0.1cm} \includegraphics[angle=0, width=12cm]{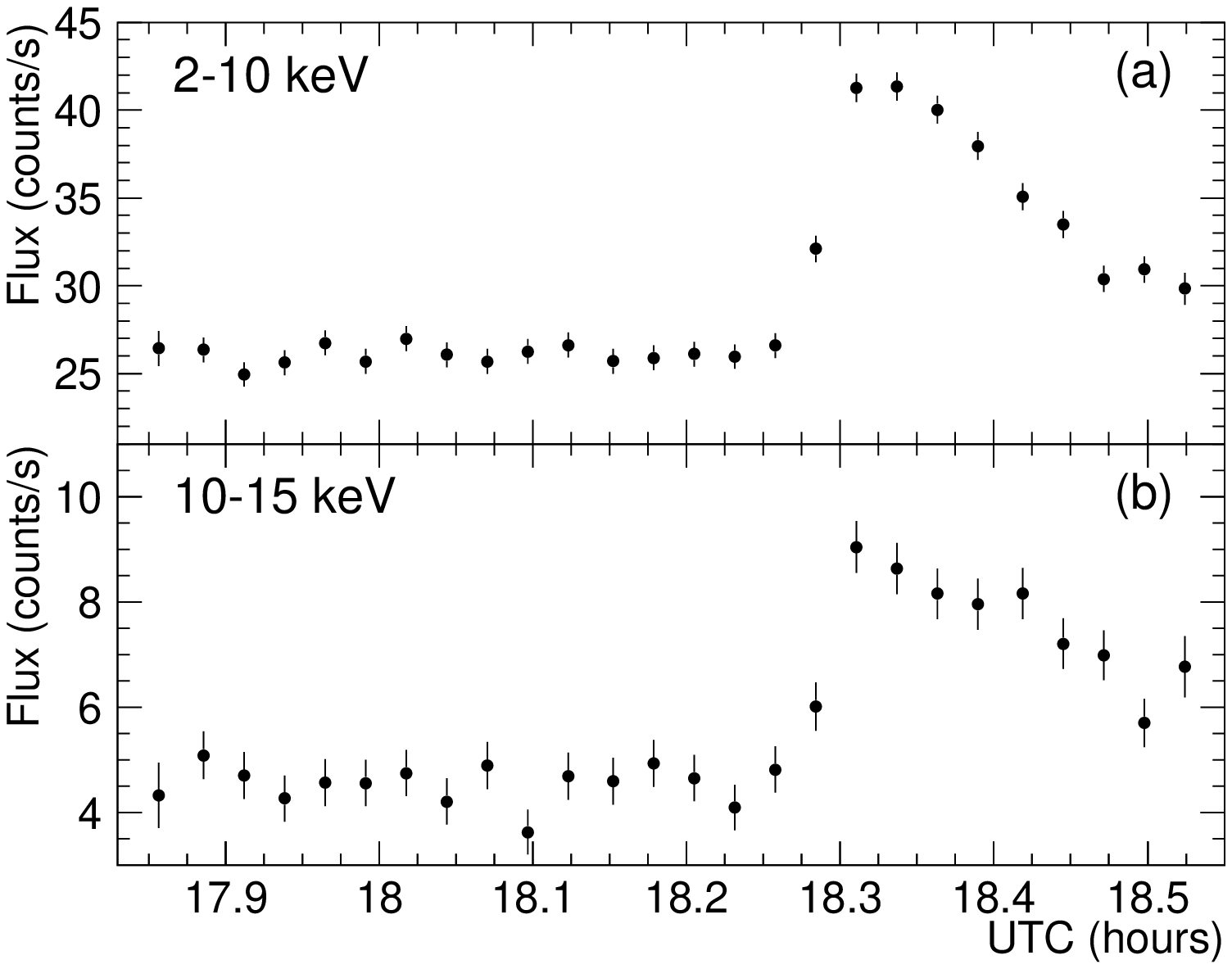}}
\vspace*{-0.5cm}
\caption{\scriptsize Lightcurves of Mkn 501 in X-rays, revealing three very different variability timescales:
 years, days, minutes.
Left: 7--day weighted averages of RXTE ASM data. Center: ASCA long observation ($\sim$10 days) 
on March 2000, 6000s bins \cite{chiharu}. The spectrum hardens during flares 
(see the different soft and hard
countrate). Right: XTE PCA observation on 25/5/98, 96s bins \cite{catanese}. 
After UTC 18.3, the spectrum hardened from $\alpha=1.02$ to $\alpha=0.87$ ($\pm0.03$).
}
%\vspace*{-0.5cm}
\label{f1}
\end{figure*}

\vspace*{-0.8cm}
\section{Variability properties}
\vspace*{-0.8cm}
Multi-wavelength observations of  variable sources require an understanding of 
their variability characteristics to decide the best observing strategy. 
The most variable and potentially strong extragalactic sources of TeV photons are Gamma-Ray Bursts
(GRBs) and blazars. 
For GRBs (whose flux decays as $\sim t^{-1}$) 
the choice is rather straightforward (albeit difficult to carry out):
to observe as quick as possible and until it fades away.

%Blazars instead 
%(and high energy peaked BL Lacs in particular, which TeV sources belong to)
Blazars instead are characterized by very different variability timescales, from years to minutes,
and not only among different objects but also in the single sources.
In high energy peaked BL Lacs (HBLs, which comprise all the detected TeV sources)
we are now aware of at least 3 main timescales, well exemplified by the Mkn 501 behaviour
in X-rays (Fig. 1):
a long-term one, from months to years (e.g. the 1997 huge flare); a ``middle-term" one, with 
a typical timescale of $\sim$1 day (as revealed by the ASCA long observations,  
also in  Mkn 421 and PKS 2155-304 \cite{chiharu}); and a short-term one, from hours down to even minutes,
as observed by RXTE  in May 1998, with a hard X-ray 
doubling timescale of $\sim$6 minutes \cite{catanese}. 
Extremely short variability is also a property of the TeV emission (e.g. Mkn 421, \cite{gaidos,aha2}),
with doubling timescales as short as 20 minutes.

\begin{figure*}[p]
\centering
\vspace*{-1.1cm}
\hspace*{-1.0cm}
\resizebox{16cm}{!}{\includegraphics[angle=0, width=15cm]{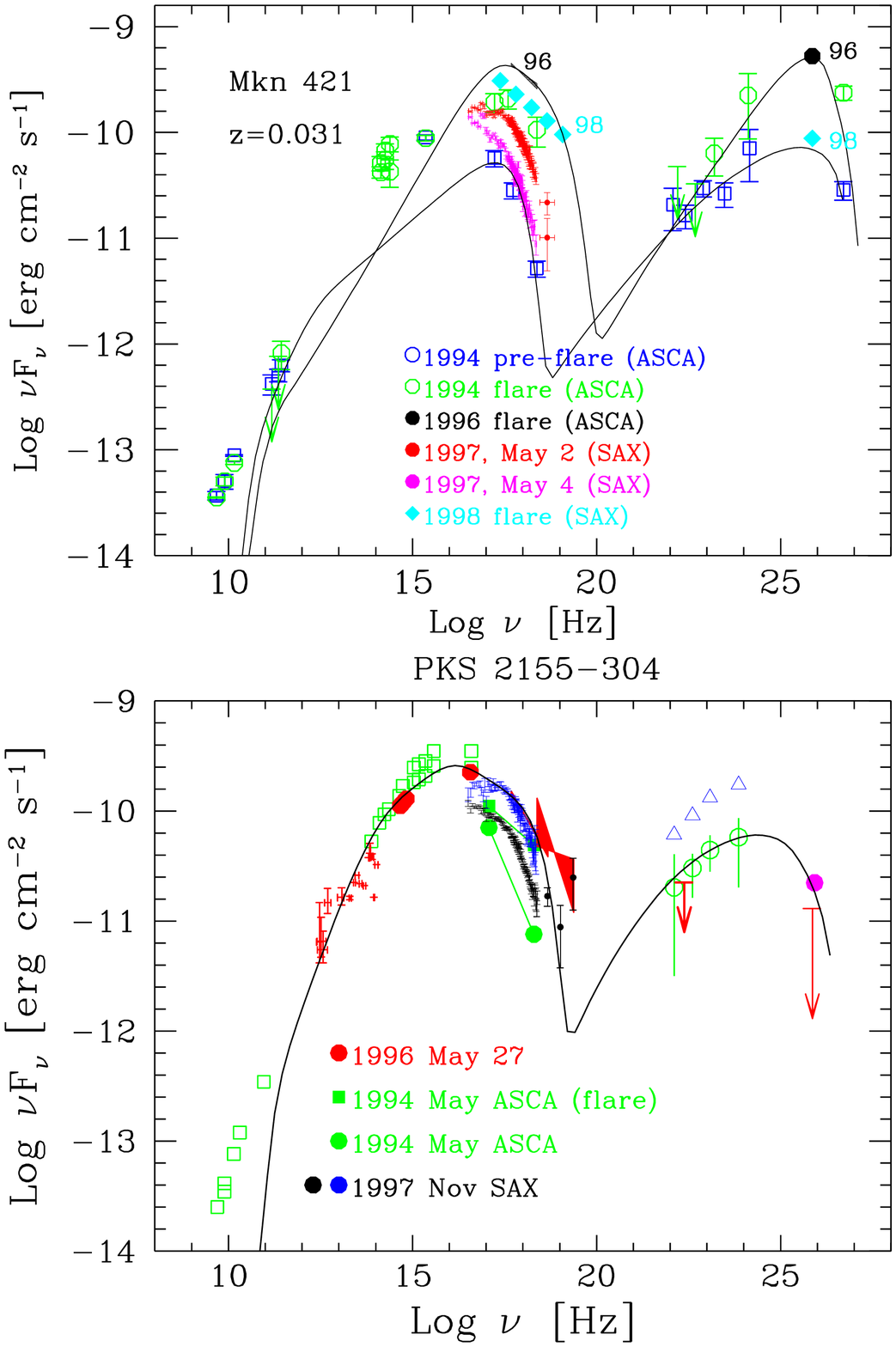}
\hspace*{-2cm} \includegraphics[angle=0, width=15cm]{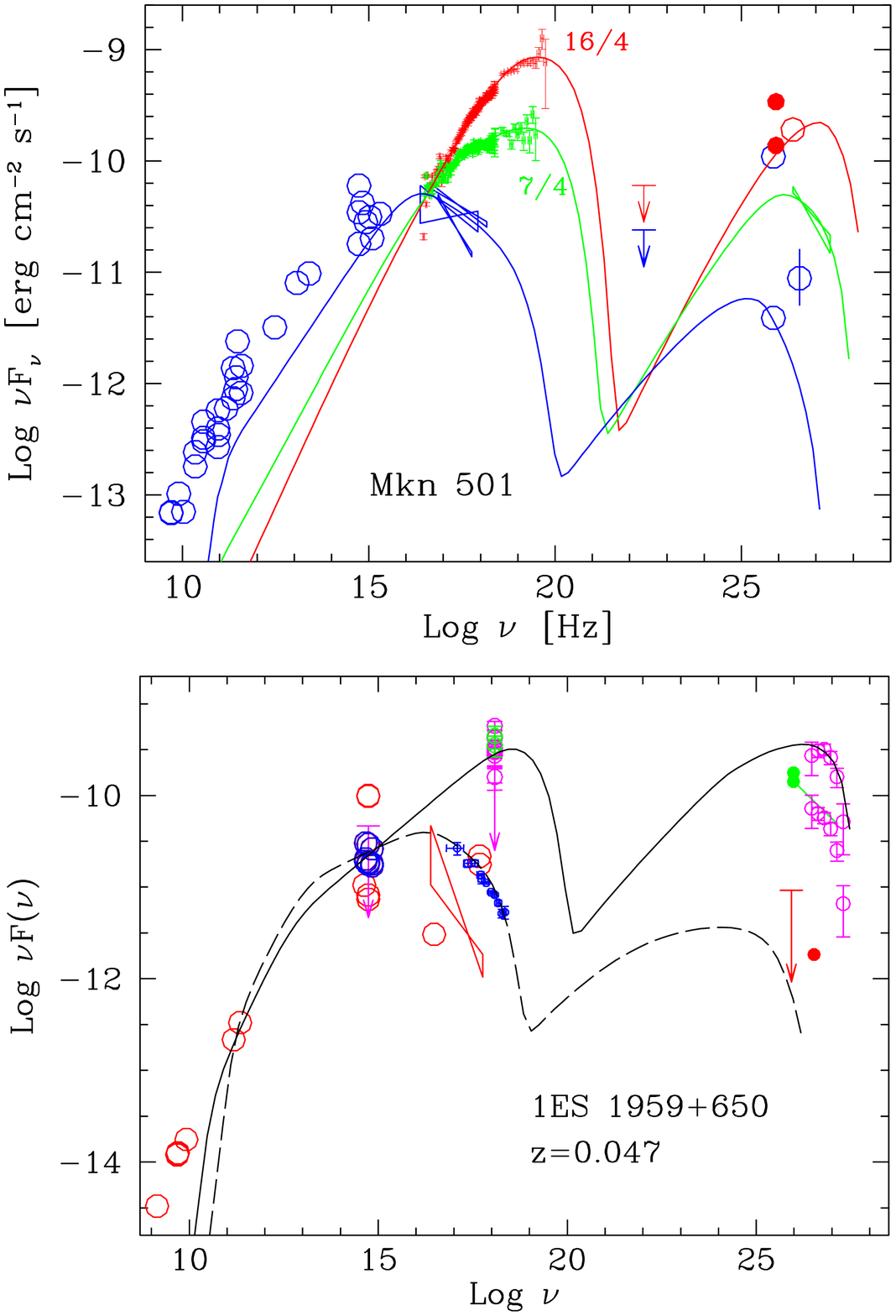}}
\vspace*{-1.5cm}
\caption{\scriptsize SEDs of four HBLs, showing two characteristic ``flaring modes" for the synchrotron peak. 
Left (Mkn 421, PKS 2155-304): as the flux varies, the peak energy remains nearly stable.
Right (Mkn 501, 1ES 1959+650): as the flux increases, the peak shifts at high energies, 
with a pivoting in the optical--soft-X band. 
For 1ES 1959+650 during the May 2002 flare, the pivoting is inferred by the high X-ray flux and hard spectrum
($\alpha\approx0.6$, \cite{henric})
and the contemporaneous optical upper limit (MERCATOR telescope; details in \cite{horn19}). 
See \cite{henric} for the results of the multi-wavelength campaign.
}
\label{f2}
%\end{figure*}

%\begin{figure*}[t]
%\centering
\vspace*{-0.8cm}
\hspace*{-0.8cm}
\resizebox{16cm}{!}{\includegraphics[angle=0,width=19cm]{1426test.epsi}
\hspace*{-0.4cm} \includegraphics[angle=0, width=13.5cm]{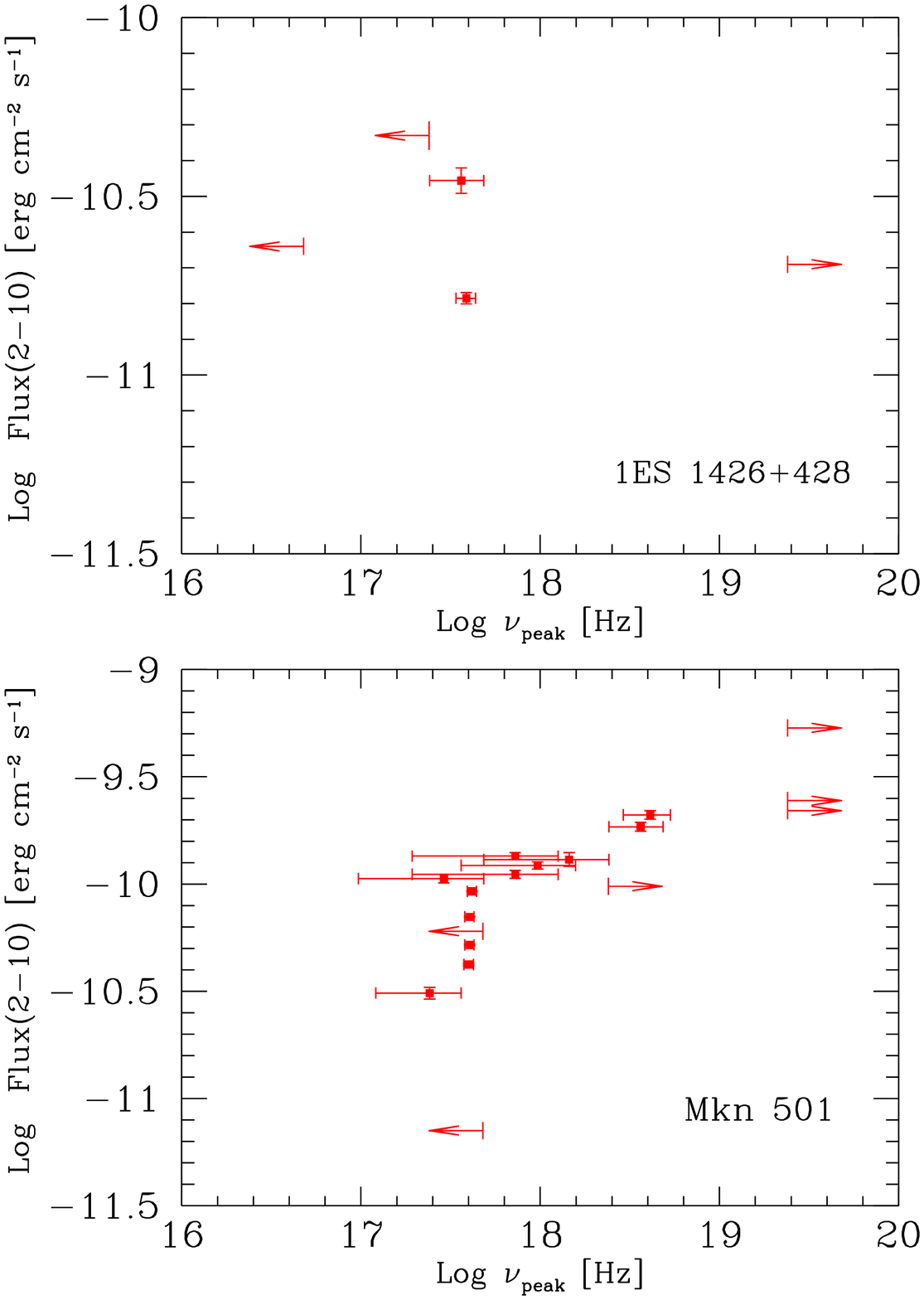}}
\vspace*{-1.5cm}
\caption{\scriptsize Left: SED of 1ES 1426+428 \cite{costa1}. The upper TeV data correspond to the absorption
corrected ones (using the Primack EBL model, \cite{costa2}). This source is 
characterized by large peak shifts not coupled with large flux
variations, as opposed to Mkn 501 (see right). 
Right: historical X-ray data. The peak frequency is determined by curved or broken power-law best fits,
with upper and lower limits given by the X-ray spectrum being steeper or harder than 1, respectively,
over the whole instrument's band.}
\label{f3}
\end{figure*}

Connected with the flux variability, HBLs show also strong spectral variations,
which often change the position of the SED peaks.  
Although there seems to be a trend (in the X-ray and TeV bands) for ``harder when brighter" spectra,
there is not a unique relation between flares and (synchrotron) peak changes.
In fact, we can outline 3 types of behaviour:  a ``Mkn421--like" one (Fig. 2 left),
where the peak frequency changes very little despite large flux variations \cite{gfoss};
a ``Mkn501--like" one (Fig. 2 right), where during flares the peak shifts towards higher energies 
by orders of magnitude (e.g from $<$1 to $\geq$100 keV); and a ``1ES 1426+428--like" one (Fig. 3), 
which seems characterized by large peak variations (of the same magnitude of Mkn501) but
during roughly constant flux states (at least in the X-ray band).
A ``Mkn501--like" behaviour has been recently followed by 1ES 1959+650 during the May 2002 
strong flare \cite{henric},
with striking similarities  also in the 'pivoting' of the SED
around the optical--soft-X range \cite{horn19}.

All these differences and similarities suggest that there are common but
intrinsically different parameters/mechanisms driving the variability, whose understanding
requires to follow the overall SED changes on all typical timescales.
%%i.e. to have simultaneous {\it spectral} data in all the 3 bands.

\vspace*{-0.6cm}
\section{The X-ray--TeV connection}
\vspace*{-0.8cm}
The interest of X-ray and TeV observations is that they sample electrons of
%simultaneous X-ray--TeV observations is that they sample electrons of 
roughly equal energy, emitting through two different processes. In HBLs, moreover,
 the bulk of the luminosity is emitted in those bands  and  it appears to be produced
by the {\sl same} electron population, as suggested by the 
often observed tight correlation between the two emissions. 
A strong confirmation  of this last hypothesis 
has come from a simultaneous observation of Mkn 421 in 1998 (Fig. 4), where
the peaks of the emissions were correlated within 
one hour (although with different halving times; \cite{maraschi}).

\begin{figure*}[t]
\centering
%\vspace*{-0.5cm}
\hspace*{-0.6cm}
\resizebox{16.5cm}{!}{\includegraphics[angle=0, width=11cm]{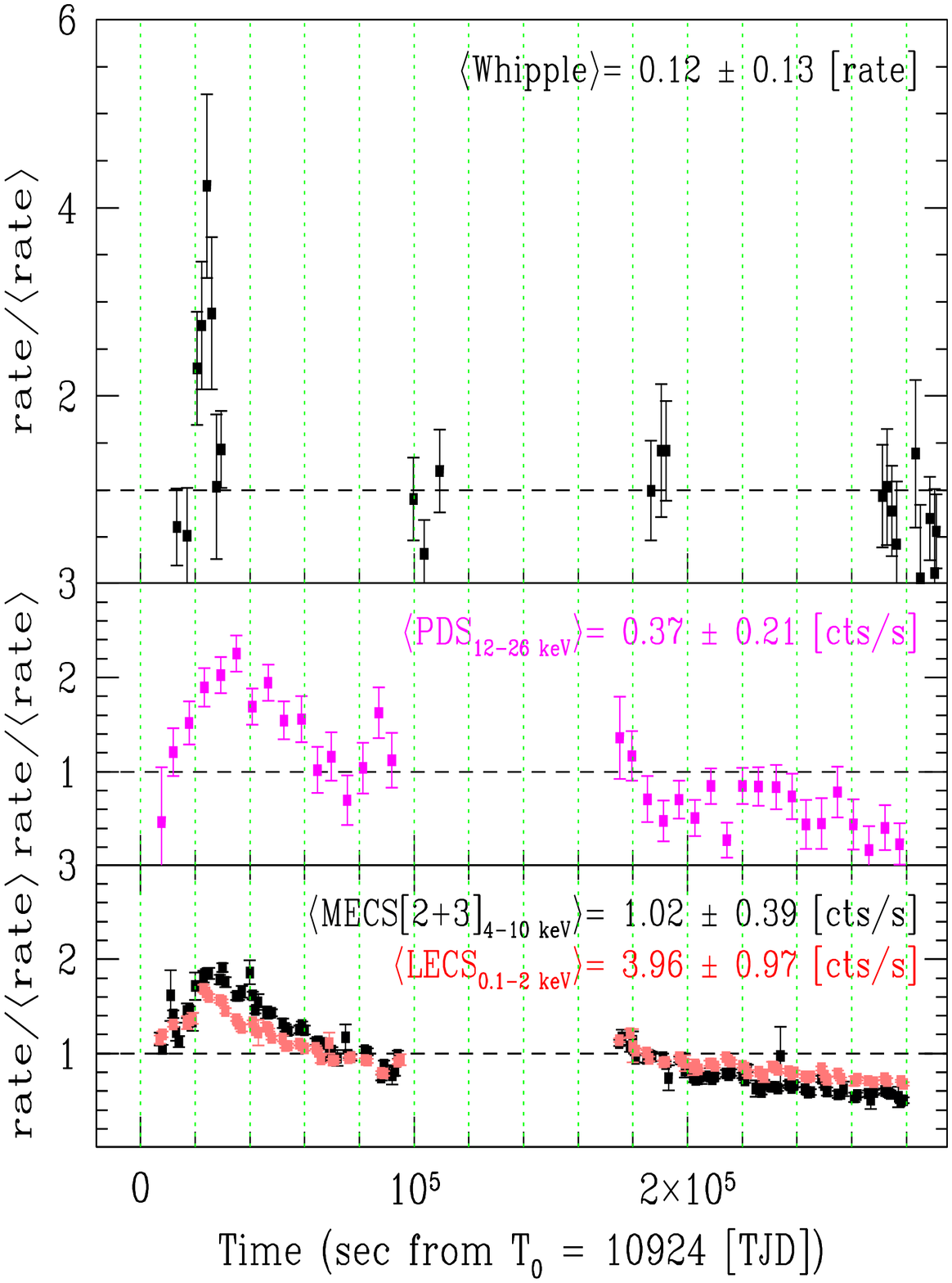}
\hspace*{-0.7cm} \includegraphics[angle=0, width=18cm]{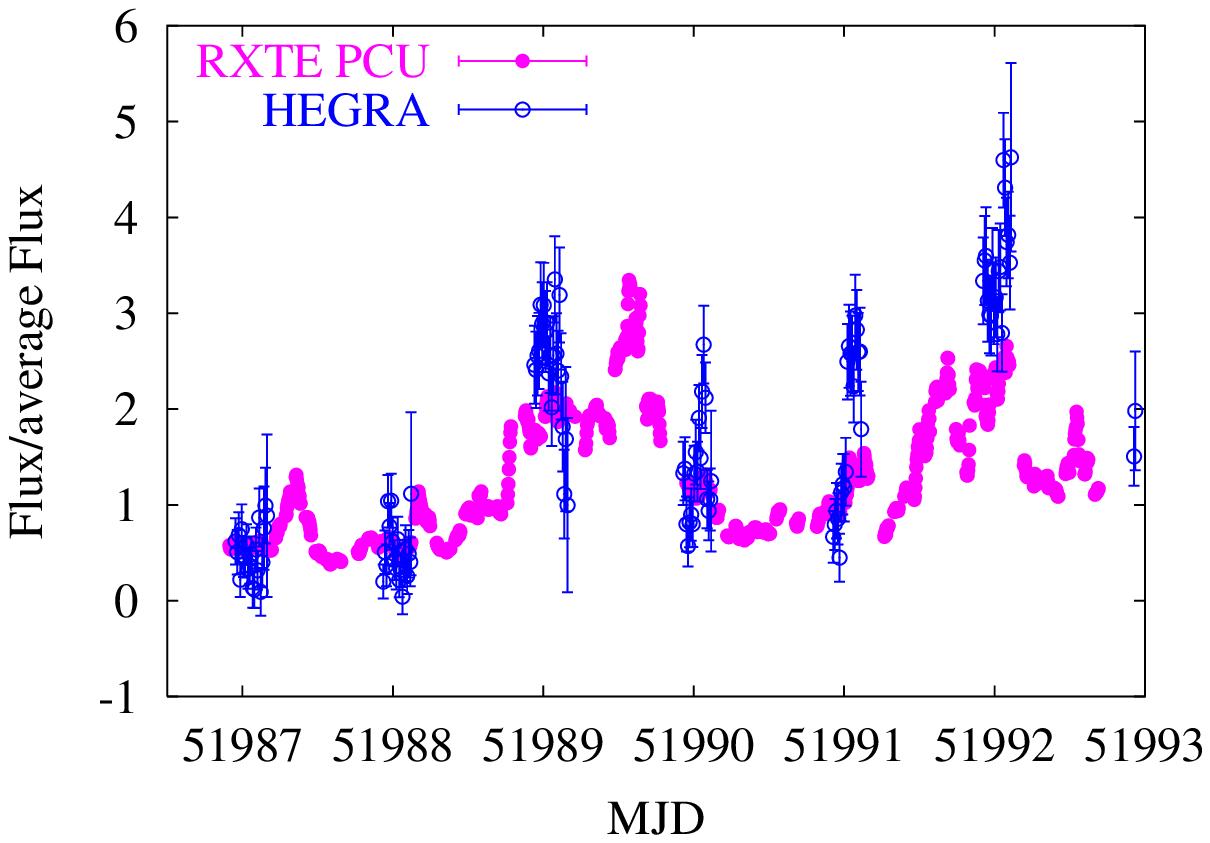}
 }
\vspace*{-1.cm}
\caption{\scriptsize Left: simultaneous lightcurves of Mkn 421 by WHIPPLE ($>2$TeV) and \sax, 
in April 1998 \cite{maraschi}. Right: a simultaneous one week monitoring of Mkn 421 by RXTE (PCA) and HEGRA, 
in March 2001 \cite{horns421}.
}
\label{f4}
\end{figure*}

However, new results are now showing that this relation may be more complex than expected, since the two bands 
does not always follow the same behaviour. 
An example is given again by Mkn 421, in 2001 (Fig. 4 right): 
although there is a general correlation, there are periods
where we see a strong decrease (MJD $\sim$51989) or a rapid flare (MJD $\sim$51990) of the TeV flux
not accompanied by relevant changes in the X-ray band. This behaviour seems also confirmed 
by 1ES 1959+650 in its 2002 active state, where evidence was found for a so called ``orphan" \gamm-ray
flare (i.e. not accompanied by an X-ray flare) \cite{henric}.    
More observations, possibly over several days, are necessary, to see if there are also
X-ray flares not accompanied by TeV flares, and to unveil 
possible lags between them and with the seed photons fluxes.

\begin{figure*}[t]
\centering
%\vspace*{-1.0cm}
\hspace*{-0.6cm}
\resizebox{16cm}{!}{\includegraphics[angle=0, width=14cm]{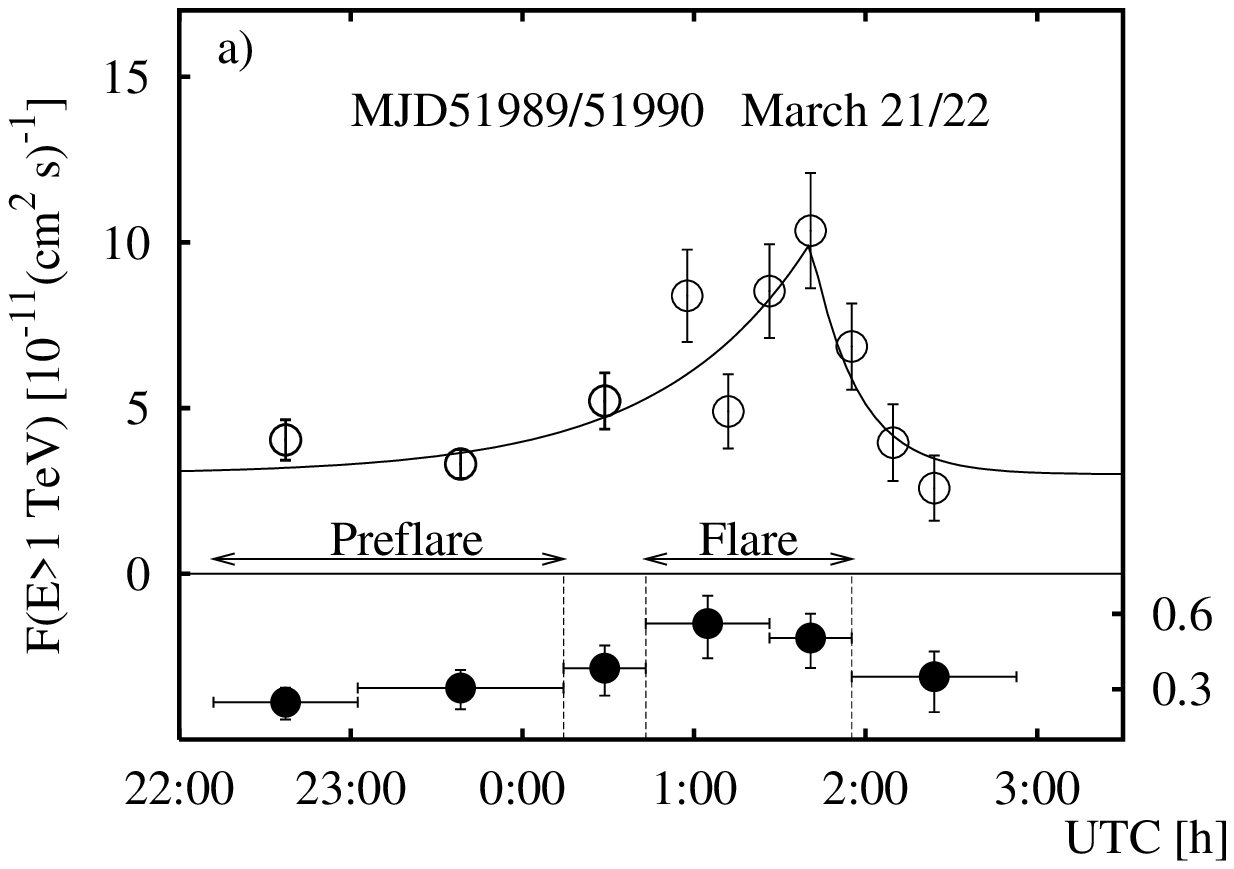}
\hspace*{-0.2cm} \includegraphics[angle=0, width=14cm]{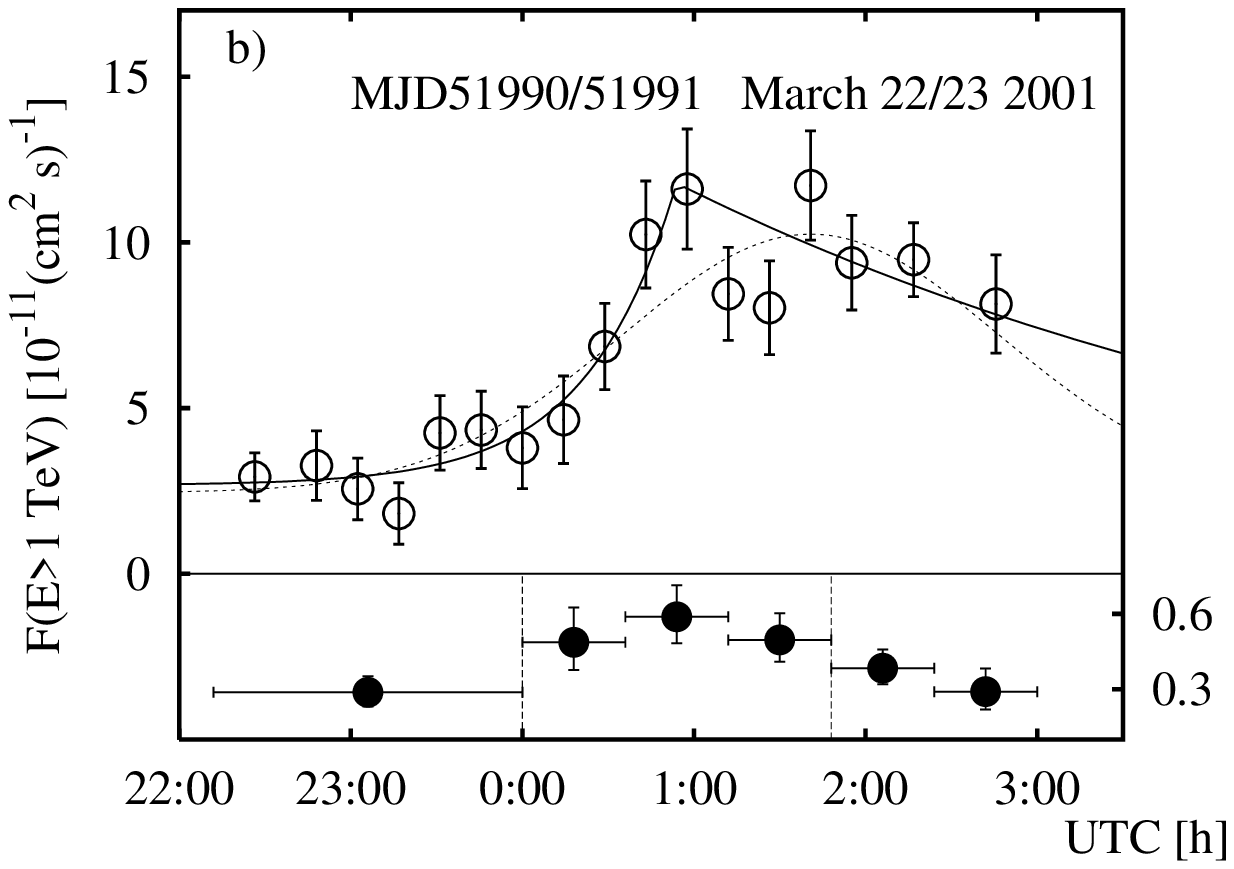}}
\vspace*{-1.0cm}
\caption{\scriptsize Two intra-night flares of Mkn 421 at TeV energies (HEGRA data, taken on two 
consecutive nights), showing different rise and decay timescales \cite{aha2}.
Below, the hardness ratio is reported
(calculated from the bands 0.75--1.5 and 1.5--4.0 TeV).  
}
\label{f5}
\end{figure*}

For both diagnostics and model testing, the intra-day activity of TeV blazars is 
perhaps the most challenging, since
the shortest TeV timescales pose the tightest constraints to the size of the emitting region and to the
\gamm-\gamm~transparency \cite{celotti}. Moreover, recent HEGRA data on Mkn 421 have 
shown an entire ``zoo" 
of flares with quite different rise and decay times (e.g. Fig. 5), hint
of a complex interplay between acceleration and cooling/escape times.
The improvement in sensitivity ($\sim$1 order of magnitude) of the new generation CT,
coupled with the continuous coverage provided by the large area X-ray telescopes (CHANDRA, XMM),
are now making possible to follow the spectral evolution in both bands on hour timescales and without
gaps, allowing to study the inner jet conditions with unprecedented detail.

%In the past such studies were difficult due to the lack of sensitivity in the TeV band,
%but now with the upcoming of a new generation of CT

Another fundamental advantage of the new generation CT is their lower energy thresholds ($\leq$150 GeV),
which will provide data also at energies not (or less) affected by absorption due 
to \gamm-\gamm~collisions with the cosmic Extragalactic Background Light (EBL, \cite{steck,primack}). 
Together with the study of the correlated X-ray--TeV variability,
from simultaneous observations we hope to be able to 
constrain the whole SED model parameters 
such to ``measure" the EBL by comparing the predicted intrinsic TeV spectrum 
with the observed one \cite{coppi, kraw}. 

\begin{figure*}[t]
\centering
\vspace*{-0.8cm}
\hspace*{-1.1cm}
\resizebox{17cm}{!}{\includegraphics[angle=0, width=17.5cm]{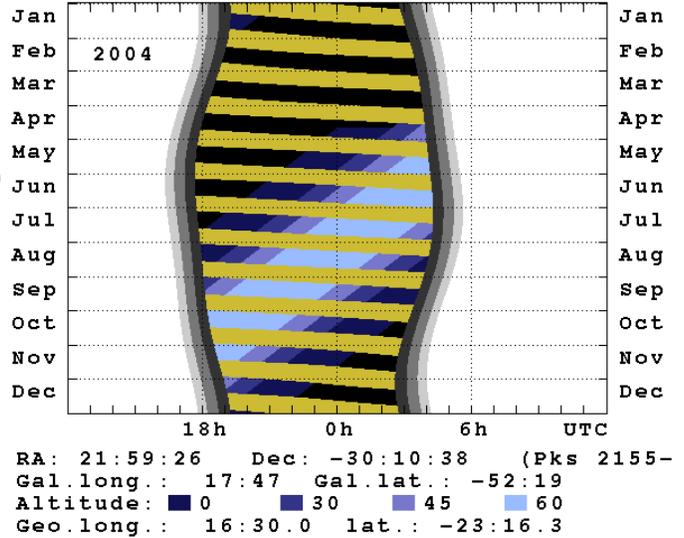}
\hspace*{-1cm} \includegraphics[angle=0, width=18cm]{2155hess.epsi}
 }
\vspace*{-1.6cm}
\caption{\scriptsize Left: sky visibility plots for XMM and CHANDRA (from the respective handbooks).
Right: visibility of PKS 2155-304 by \hess during 2004, above a certain altitude (in degrees).
The nearly horizontal stripes correspond to the moonlight--contaminated nights, i.e. not usable by CTs.
For simultaneous observations, while the CHANDRA window goes from April to end December, 
the XMM one is limited to  28/4--27/5 and 31/10--26/11.
}
\label{f6}
\end{figure*}

\vspace*{-0.8cm}
\section{Observation strategies}
\vspace*{-0.8cm}
It is clear at this point that we need simultaneous observations in all the three main bands
{\it with spectral information}. Spectra are precious 
also in the ``optical" band, 
to outline the overall SED shape (and peak position) and to follow its evolution during flares. 
The ``optical" band is important also because changes in the seed photons flux or/and spectrum
may give origin to different behaviours between the TeV and X-ray emissions.

%be a factor to explain at the basis of different behaviours of the TeV and X-ray emissions.

The 'long-term' variability requires of course a long term monitoring program of a number
of objects (as now done by RXTE ASM), which should continue to be done preferably at X-rays, 
since an only optical
monitor would have missed the ``Mkn501-like" flares, due to the pivoting of the SED at low frequencies.
Simultaneous campaigns over several days, maximizing the continuous coverage through the coordination
of ground CT and optical telescopes, 
will allow instead to sample both the middle and short-term variabilities,
and to reveal possible lags.
Long observations will allow also to investigate  the relations between these 
two variabilities, e.g. if the shortest one appears only on top of the other or not,
and if with different characteristics.
For the same reason, simultaneous campaigns should be performed not only 
during the flares detected by the long term monitoring, 
but also in quiescent periods.

\vspace*{-0.8cm}
\section{A warning about X-ray coverage}
\vspace*{-0.8cm}
Together with CHANDRA, XMM is in principle the best X-ray instrument for this type of studies,
thanks to its large collecting area (providing spectra on shorter timescales)
and its long orbital period, which allows for continuous observations without gaps up to $\sim$37 hrs.
Regrettably, the strong constraints on the solar panels orientation (the sun avoidance angle
has to be $70^{\circ}\leq\theta\leq110^{\circ}$) severely affects the satellite performance
in two for us important areas: the ability to perform observations triggered by flaring states
and simultaneously with ground telescopes.
As shown in Fig. 6 (left), the visibility of an object during the year  is limited for most of the sky
to $\sim$2-3 months, with correspondingly low probability that a flare will occur 
in the right epoch. Even more important, the constraints involve also the anti-solar direction
(contrary to CHANDRA), thus making impossible to observe a source when it is visible under the best conditions 
from ground telescopes (i.e. for the full night).

An example is given by PKS 2155-304 with \hess (Fig. 6, right):
%, assuming to plan a simultaneous X-ray--TeV observation with \hess: 
the source can be pointed by XMM only in May
(28/4--27/5) and November (31/10--26/11), i.e. when it can be observed by \hess for only $\sim2$ hours
at the end or beginning of the night. 
The same problems affect also INTEGRAL, although with slightly less constraints on the sun angle and thanks to the
wide field of view of its instruments (JEM-X f.o.v. is 4.8$^{\circ}$, fully coded).
Since SWIFT will be devoted mainly to GRBs studies (but see \cite{gehrels} for its possible use also 
with flaring blazars),
this leaves the TeV community for the next years (i.e. when the new generation CTs are becoming operative)
with only CHANDRA and RXTE as the main X-ray satellites for simultaneous campaigns.  

%So at present, with the upcoming of the new generation CT, we have only Chandra with the area and continuum
%visibility characteristics fo a seriuos investigation of variable, not predictable phenomena;

%\subsection{Another sub section}
%\subsubsection{A subsub section} with some text in it.
%\paragraph{And a paragraph}
%(ie a sub sub subsection)

\end{document}